\documentclass{vgtc}                          

\graphicspath{{figures/}{pictures/}{images/}{./}} 

\usepackage[most]{tcolorbox}
\usepackage{xcolor}
\usepackage{tikz}
\usepackage{times} 
\usepackage{wrapfig}
\usepackage{enumitem}
\usepackage{cite}

\usepackage{tabu}                      
\usepackage{booktabs}                  
\usepackage{lipsum}                    
\usepackage{mwe}                       

\usepackage{mathptmx}                  

\usepackage{subcaption}
\onlineid{1232}

\vgtccategory{Research}

\vgtcinsertpkg

\title{Made to Feel: How Designers Bring Emotions into Affective Visualization}

\author{%
\vspace{-10pt}
Yixin Bai\footnotemark[1] 
\and Ziyi Wang\footnotemark[1] 
\and Keke Wu\footnotemark[1] \and Fumeng Yang\footnotemark[1]\thanks{e-mail:\{yixbai01, zoewang, kekewu, fy\}@umd.edu\vspace*{-35pt}}
}
\affiliation{\vspace{-8pt}\scriptsize University of Maryland, College Park
\vspace{-5pt}
}

\teaser{
  \centering
    \vspace{-8pt}
\includegraphics[width=\linewidth]{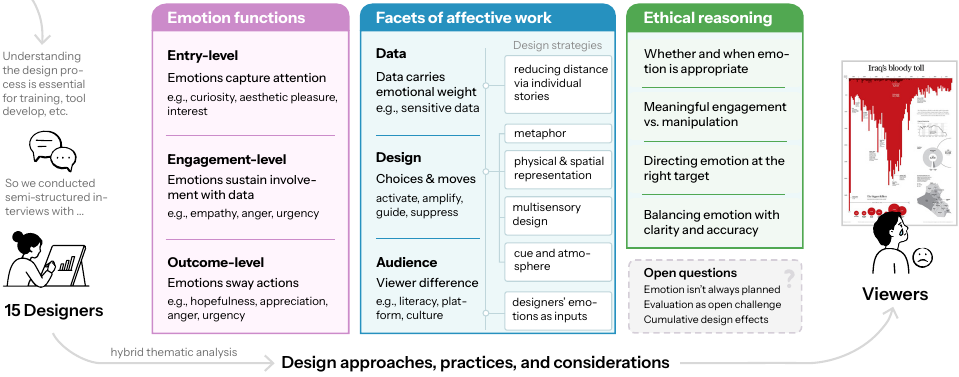} 
    \vspace{-16pt}
  \caption{%
  \textbf{Overview of our method and findings.} We conducted semi-structured interviews with 15 visualization designers. From the interviews, we identify the functions of emotion for viewers, the facets of how designers work with emotions, and the ethical considerations in designing affective visualization. Icons were created using GPT Image 2 or under a Flaticon license. 
  }
    \label{fig:teaser}
}

\usepackage[normalem]{ulem}
\newif\ifnotes
\notestrue 

\newcommand{\px}[1] {P#1}

\definecolor{editcolor}{HTML}{fc035a}

\definecolor{fycolor}{HTML}{8654d1}

\definecolor{keke}{HTML}{4a90e2}

\usepackage{soul}

\newcommand{\etal}{\mbox{et al.}\xspace}
\definecolor{DesignColor}{HTML}{2691bf}
\definecolor{BlackColor}{HTML}{000000}
\definecolor{NavyBlue}{HTML}{000000}
\definecolor{ethicsColor}{HTML}{499949}
\newcommand{\ethics}[1]{\textcolor{ethicsColor}{#1}}
\definecolor{threeColor}{HTML}{2691bf}

\definecolor{RoleColor}{HTML}{bc69b6}
\newcommand{\roleText}[1]{\textcolor{RoleColor}{#1}}
\newcommand{\designText}[1]{\textcolor{DesignColor}{#1}}
\newcommand{\designUL}[1]{%
  \begingroup
  \setulcolor{DesignColor}%
  \textcolor{BlackColor}{{\fontsize{8.5pt}{8.5pt}\selectfont\textit{\ul{#1}}}}%
  \endgroup
}
\newcommand{\ethicsUL}[1]{%
  \begingroup
  \setulcolor{ethicsColor}%
  \textcolor{BlackColor}{\textit{\ul{#1}}}%
  \endgroup
}
\newcommand{\designIB}[1]{%
  \begingroup
  \setulcolor{DesignColor}%
  \textcolor{DesignColor}{\textit{\textbf{#1}}}%
  \endgroup
}

\definecolor{revcolor}{HTML}{E67E22}   

\newcommand{\eg}{\mbox{e.g.,}\xspace\@}

\newcommand{\inlinequote}[1]{\textit{``#1''}}

\usepackage{fdsymbol}
\newcommand{\heart}{\ensuremath\heartsuit}

\makeatletter

\renewcommand\paragraph[1]{%
  \@startsection{paragraph}{4}{\z@}%
  {.3em \@plus .2em \@minus.0em}%
  {-.5em}%
  {\normalfont\sf\bfseries\fontsize{8.25pt}{8.25pt}\selectfont}%
  {#1}%
}
\makeatother

\abstract{%
    Affective visualization is increasingly studied in visualization research, yet how designers bring emotions into their visualization work remains unexplored.
    This paper addresses this gap through semi-structured interviews with 15 visualization practitioners. Using hybrid thematic analysis, we identify: 
    (1) three functions that emotions can serve for viewers (entry, engagement, outcome); 
    (2) three facets of how designers work with emotion (data, design, audience), along with design strategies; and (3) ethical considerations in the design process. 
    We also observe that affective intent often emerges during the design process rather than being planned from the outset, and that emotional impact arises from accumulated design choices rather than isolated visual elements. Finally, we highlight evaluation as a key challenge identified by our participants. 
    } 


\begin{document}



\firstsection{Introduction}

\maketitle




Emotion is a fundamental aspect of human decision-making and behavior~\cite{Damasio:1994:DES}. Data visualization, meanwhile, is typically designed to communicate complex information visually, supporting cognitive processes and informed decision-making~\cite{Card:1999:RIV,Munzner:2014:VAD, Kale:2020:VRS}.
~At the intersection of these two concepts lies \textit{affective visualization}, visualization intentionally designed to evoke, communicate, or influence emotion.
Consider Simon Scarr's \textit{Iraq's Bloody Toll} (\cref{fig:teaser} right end)~\cite{Scarr:2011:Iraq}, which depicts coalition casualties from the U.S. military campaign in Iraq. 
The inverted red bars conjure a visceral metaphor of bloodshed, leaving viewers with a sense of anger, grief, or sympathy. 
Beyond aesthetics, such affective responses can go on to influence viewers' attitudes and behaviors~\cite{Yang:2024:SwayPublic}.

Prior research has primarily approached affective visualization either by analyzing completed designs~\cite{Lan:2024:AVD, Prantl:2024:URE} or by measuring viewer responses, such as how visual elements shape arousal and valence~\cite{Blair:2024:QER, Boy:2017:SPB}. While these works offer valuable insights into how affective visualizations are perceived and experienced, the design process itself remains underexplored. In addition, emotion is a powerful means of influence, and how designers navigate the boundary between engaging and manipulating viewers raises ethical questions that are equally unexamined. Understanding the design process is therefore essential for the field to move forward. It can reveal the tacit knowledge, recurring challenges, and practical workflows that provide a foundation for designer education, tool development, and guidelines for intentional affective design.

\looseness=-10

We address this gap through a qualitative study with visualization designers. We conducted semi-structured interviews with 15 practitioners who had each produced at least one publicly available visualization project that we considered affective (\cref{sec:method}). Through hybrid thematic analysis on 866 codes, we identify (\cref{sec:findings}):
(1) three levels of emotional functions (entry, engagement, outcome); 
(2) three facets of how designers bring emotion into the design process (data as material, design as mediation, audience as calibration), along with corresponding design strategies; and
(3) ethical considerations that arise throughout the design process. 
We also highlight evaluation as a key challenge, and propose a forward direction for evaluation methods
(\cref{sec:discussion}).

These findings on designers' approaches, practices, and considerations form the contributions of this work.
While qualitative in nature, they advance our understanding of affective design process and lay the groundwork for future evaluation methods and practical guidance for designers and researchers who work with emotion. Our supplementary materials are available at {\small\texttt{\href{https://doi.org/10.17605/OSF.IO/RFVZU}{10.17605/OSF.IO/RFVZU}}}. 


\section{Related Work \& Background}

Affective visualization is defined as visualization ``designed to communicate, or influence emotion''~\cite{Lan:2024:AVD}.
It overlaps with several other visualization techniques but is distinguished by its focus on emotion. 
For example, narrative visualization~\cite{Segel:2010:NVT} uses storytelling structures such as the martini glass or drill-down story, where emotion may arise from effective storytelling but is not the design focus. 
Data Humanism~\cite{Lupi:2017:DHR, Alhazwani:2025:DHD} advocates treating data as human-made, imperfect, contextual, and personal, and emotion as a byproduct of humanizing data. 
Affective visualization, in contrast, targets emotion, regardless of whether the data is presented in a humanized form or organized as a narrative.


Prior research has analyzed existing affective visualizations as finished artifacts to surface design patterns. Lan et al.~\cite{Lan:2022:NEP} investigated negative emotions in serious data stories, finding gaps between expert-proposed methods and viewers' actual responses. Lan~\etal~\cite{Lan:2025:MWF} further examined affective geovisualization design through the lens of relationships between people and place, identifying how designers translate emotional connections to place into visual form. Bartram~\etal~\cite{Bartram:2017:ACV} demonstrated that color systematically influences emotional associations, though these associations are not culturally universal~\cite{Prantl:2024:URE}. 
Beyond identifying patterns, other studies have built on artifact analysis to develop frameworks or test whether intended responses match viewer experience~\cite{LeeRobbins:2022:ALO,LeeRobbins:2026:AAR}. 

Other empirical work has measured how viewers' affective responses vary with data and visualization elements. Blair et al.~\cite{Blair:2024:QER} showed that even data with no real-world meaning evokes emotion: structural properties such as trend direction, density, and variance independently affect viewers' \textbf{arousal} (i.e., how activated viewers feel)
and \textbf{valence} (i.e., whether they feel positively or negatively). 
Slovic~\cite{Slovic:2007:ILM} documented psychic numbing, where compassion decreases as the number of victims increases, and V\"astfj\"all et al.~\cite{Vastfjall:2014:CFA} found this fading begins from just one person to two. Boy et al.~\cite{Boy:2017:SPB} and Morais et al.~\cite{Morais:2021:CAP} found that anthropomorphic icons had minimal effects on empathy and prosocial behavior. 
Yang et al.~\cite{Yang:2024:SwayPublic} tested affective responses to four election forecast visualizations in a longitudinal experiment. 

The cited studies above primarily capture viewers' responses.
Recent visualization research on design processes~\cite{Parsons:2022:UDV}, inspiration practices~\cite{Baigelenov:2025:HVD}, designers' perspectives~\cite{Schuster:2026:PPD}, and decision-making under constraints~\cite{Zhang:2023:VDP} show that practitioners hold design knowledge not captured through artifact analysis or controlled experiments alone. Therefore, our study also focuses on practitioners' design process and considerations to learn their first-hand knowledge.

\section{Method}
\label{sec:method}


To understand how designers bring emotion into their own visualization designs, we utilize semi-structured interviews  and  formulate three research questions (RQs): 

\begin{itemize}[itemsep=-4pt,topsep=0pt, label={}, leftmargin=24pt]
\item[RQ1.] How do visualization designers conceptualize emotion in their design process?
\item[RQ2.] What strategies do designers use to evoke or manage viewers' emotions?
\item[RQ3.] How do designers reason about the ethical boundaries between meaningful engagement
and manipulation?
\end{itemize}

\enlargethispage{.5\baselineskip}

\paragraph{Participants.}

We required participants to have at least one publicly available affective visualization project. 
We proactively contacted potential participants through (1) reviewing visualization portfolios on platforms such as the Information is Beautiful Awards~(2019, 2022, and 2023 editions)~\cite{InfoBea}; (2) using the example corpora compiled by Lan et al.~\cite{Lan:2022:NEP, Lan:2024:AVD}; and (3) distributing a recruitment flyer through the Data Visualization Society Slack and relevant subreddits (i.e., \texttt{r/datavisualization} and \texttt{r/hci}). 
We contacted over 120 practitioners in total and asked them to share examples of their visualization work. One author reviewed all examples to select those projects that appeared to engage with emotion (the definition of affective visualization~\cite{Lan:2024:AVD}). 
We ultimately interviewed 15 qualified participants (see \cref{tab:participants}), and most of their projects were static designs.
 This study was approved by our IRB office, and each participant received a \$50 electronic gift card as compensation.\looseness=-10

\begin{table}[t]
\caption{\textbf{Participants Demographics:} Jobs, geographic regions, years of experience (YoE), and visualization projects discussed.}  
\vspace{-7pt}
\label{tab:participants}
\fontsize{6.5pt}{7.5pt}\selectfont
\centering
\begin{tabular}{
p{0.2cm}   
p{2.2cm}   
p{1.1cm}   
p{.5cm}   
p{2.3cm}     
}
\toprule
ID & Job Role & Location & YoE & Project Topic \\
\midrule
P1  & Product Designer        & USA          & 4--6  & Behavioral Economics \\
\px{2}  & Accessibility Designer  & USA          & 10+   & Dementia Map \\ 
P3  & Information Designer    & Asia   & 4--6  & Pi \& Privacy \\
P4  & Freelancer              & USA          & 7--10 & Personal Caregiving \\ 
P5  & Data Vis. Designer      & Europe       & 4--6  & Digital Transformation\\
P6  & UI/UX Designer          & Europe       & 4--6  & Electricity Consumption \\
P7  & Graphic Designer        & UK           & 7--10 & War History \\ 
P8  & Interaction Designer    & Europe       & 10+   & Environmental Change \\ 
P9  & Data Vis. Consultant    & USA          & 1--3  & Education \\ 
P10 & Software Engineer       & USA          & 10+   & Wealth Inequality\\
P11 & Researcher \& Educator  & Europe       & 10+   & Sexual Harassment \\ 
P12 & Data Vis. Designer      & Middle East  & 10+   & Sustainability \\ 
P13 & Interaction Designer    & Europe       & 10+   & COVID Contact Tracing \\ 
P14 & Data Vis. Professional  & Europe       & 10+   & Ecological Consumption \\  
P15 & Data Vis. Researcher    & USA          & 10+   & Climate Change \\ 
\bottomrule
\end{tabular}
\vspace{-15pt}
\end{table}


\paragraph{Procedure.}

Interviews were conducted remotely via Zoom by the first author (48--72 minutes, averaging 57 minutes). All interviews were initially transcribed by Zoom transcription, followed by manual error correction. 
The interview covered three topics: (1) designer intent and intuition, (e.g., what kind of feeling or reaction they were hoping viewers would have);  (2) their design strategies and workflow, (e.g., how and when emotion entered the design process), and how they iterated during a project; and (3) ethics and boundaries, (e.g., where designers draw the line between meaningful engagement and manipulation). To preserve anonymity, we used GPT Image 2 to create stylized versions of participants' projects and masked out text. We also confirmed quotes with participants and obtained the written consent to release these images and quotes. 
The interview protocol and transferred images are provided in the supplementary materials.\looseness=-10

\paragraph{Qualitative Analysis.}

We performed a hybrid thematic analysis~\cite{Fereday:2006:DRU, Swain:2018:HAT}, combining deductive and inductive coding so that we could apply concepts from existing theory while remaining open to themes emerging from the data.
The first author coded all transcripts; a second coder independently assigned extracted quotes to categories (Cohen's $\kappa = 0.97$). The two coders then discussed all disagreements until reaching a consensus. This process produced 866 total codes, with each participant contributing 37 to 90 codes. The codebook is included in the supplementary materials.\looseness=-10 

\section{Findings}
\label{sec:findings}





To begin, we first describe what participants consider ``emotion,'' organized by arousal~\cite{Russell:1980:CMA}.  
Seven participants 
targeted high-arousal emotions: \textit{empathy} (\px{9}, \px{11}), \textit{urgency} (\px{14}), \textit{anger} and \textit{claustrophobia} (\px{10}), \textit{horror} (\px{7}), \textit{awe and sympathy} (\px{4}), and \textit{outrage} (\px{15}). 
Five participants (\px{1}, \px{3}, \px{5}, \px{8}, \px{12}) 
targeted low-arousal responses: \textit{curiosity}, \textit{hope}, and \textit{aesthetic pleasure}. 
The remaining three (\px{2}, \px{6}, \px{13}) claimed that they excluded emotion in favor of clarity and trust, even though their projects appeared affective (\eg emotions like \textit{alarm}, \textit{interest}). 

\enlargethispage{.5\baselineskip}
\subsection*{Finding 1: Levels of Emotional Function (RQ1)}
\label{sec:rq1}

Designers believe emotions serve different functions at different points in 
 viewer experience, and we identify three levels: \roleText{entry-level}, 
\roleText{engagement-level}, and \roleText{outcome-level}. 
These levels are not strictly sequential. Entry and engagement emotions can co-occur, but outcome-level emotions often emerge later in the viewing experience. This mirrors the dual-process model of aesthetic experience, where initial pleasure gives way to sustained engagement~\cite{Graf:2015:DPP}. 
\looseness=-12


\paragraph{\roleText{Entry-level.}} 
When viewers first encounter the visualization, five participants aim to evoke an initial emotional response.
These are low-arousal emotions such as curiosity and aesthetic pleasure. The emotion does not need to be deep or specific, but must hold viewers' attention long enough for them to look closer before engaging the visualization. 
P8 attracted viewers through something \inlinequote{colorful and nice,} P1 used a deliberately \inlinequote{fun, interesting approach,} and P5 wanted to \inlinequote{make others feel curious.}

\paragraph{\roleText{Engagement-level.}} 
Once viewers attend to the visualization, seven participants try to engage them by evoking high-arousal emotions such as empathy, anger, and urgency that drive viewers to process the data deeply and connect it to their own experience. Unlike at the entry level, the type and direction of emotion matter here, as P15 put it, \inlinequote{not just whether viewers feel something, but whether they feel the right thing in the right direction.} 

\paragraph{\roleText{Outcome-level.}} 
At the outcome level, emotion is a 
means to a broader goal, shifting from what viewers 
feel to what they do, believe, or understand as a result. Nine participants described both the emotions viewers should carry away and the goals those emotions serve. P12 wanted viewers to feel \inlinequote{hopefulness for the future,} while P4
wanted them to feel \inlinequote{awe, appreciation, and sympathy,} and P1 hoped the visualization would \inlinequote{trigger their own thinking.} The goals include awareness (\px{11}: \inlinequote{bring\textrm{[}s\textrm{]} people to become aware of the phenomenon}), solidarity (\px{4} wanting mothers to feel \inlinequote{seen, that we were going through this hard thing together}), self-assessment (\px{11}: prompting viewers' \inlinequote{self-assessment of the way they behave}), and action (\px{14}: \inlinequote{People should come into action}). 
Anger and urgency may favor action, while hope sustains long-term impact on behavior~\cite{Nabi:2003:EFE,Nabi:2018:FCC}. 

\subsection*{Finding 2: Three Facets of Affective Work (RQ2)}
\label{sec:rq2}

We identify three facets of how designers work with emotion: the \designText{data} as affective material, the \designText{design} as affective mediation, and the \designText{audience} as affective calibration.
Within each facet, we also describe the specific design strategies mentioned by participants. 


\paragraph{\designText{Data as affective material.}} 
Some data carries inherent emotional weight independent of any design choices.
P11 described working with testimonies of harassment as \inlinequote{data that carries emotional burden,} and P7's war casualty data was similarly sensitive. 
By contrast, others, such as P14's energy consumption statistics and P3's Pi visualization, were not. A key strategy for this facet is to \designIB{reduce distance through individual stories}. 
Five participants resisted aggregation to maintain emotional proximity. P11 argued that \inlinequote{if you start to extract statistics, it's not working,} and that \inlinequote{numbers were against our goal.} P9 designed individual journey visualizations for eight people instead of aggregate statistics. This pattern resonates with psychic numbing, where compassion decreases as victim numbers rise~\cite{Slovic:2007:ILM, Vastfjall:2014:CFA}.

%
%
%


\setlength{\columnsep}{6pt}
\begin{wrapfigure}[15]{r}{88pt}
\vspace{-2.2\baselineskip}
\includegraphics[width=88pt]{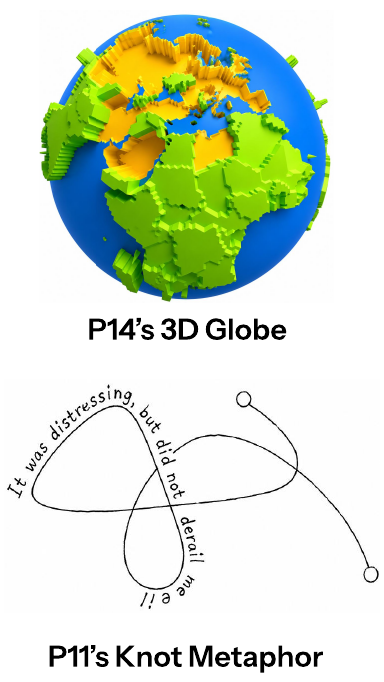} 
\vspace{-1.5\baselineskip}
\end{wrapfigure}


\paragraph{\designText{Design as affective mediation.}} 
Designers described design both as a set of visual elements, and as a way to mediate how emotion emerges from data and reaches viewers. Across projects, this mediation involved four recurring moves.
Designers may \designUL{activate} emotions by bringing them out from data: P14's 3D globe revealed countries \inlinequote{eating up the Earth} in a way the bar chart could not. Designers can also \designUL{amplify} emotion that already exists in the data. P11 used a metaphor, individual representation, and audio for already-sensitive data, explaining that \inlinequote{in this way, the emotional impact is increased.} Designers \designUL{guide} emotion toward specific targets. P9 directed viewers toward empathy rather than pity, and P15 distinguished calibrated from miscalibrated outrage. Finally, designers may \designUL{suppress} emotion. P2 designed a dementia map to be \inlinequote{objective,} explaining: \inlinequote{I think my design is objective. It's not going to rile up people's emotions.}

\enlargethispage{.5\baselineskip}
Within this facet,  four design strategies recur across projects.
\begin{itemize}[itemsep=-4pt,topsep=-2pt,label={\small\designText{$\heart$}},leftmargin=10pt]
    \item \designIB{Using metaphor}. Five participants used this strategy to connect abstract data to something viewers could feel. P11 used a knot metaphor because, in some stories, people \inlinequote{described feeling like they had a knot in their stomach.} Other metaphors were invented by designers. P14's globe with countries \inlinequote{digging into the Earth} was a metaphor for overconsumption. Still others drew on shared cultural meaning, as when P7 incorporated the poppy as a symbol tied to the aftermath of world wars.
    \item \designIB{Using physical and spatial representation}. Five participants used layout and spacing to create physical, felt responses. P10 stated that \inlinequote{the power of data visualization is that we're really good at seeing patterns and feeling space,} using stick figures packed into a tiny space to make wealth inequality felt. 
    P8 designed data as \inlinequote{digital furniture,} an object that occupies domestic space and is encountered repeatedly over time. This aligns with Blair et al.'s~\cite{Blair:2024:QER} finding that visual density affects arousal.
    \item \designIB{Considering multisensory design.} Three participants extended visual channels to auditory channels. P11 added audio readings of harassment excerpts \inlinequote{to break the visual barrier and create a more immersive environment.} P5 chose background music that \inlinequote{can create a little bit of feelings or sentiments when you listen to it.} Sound is known to trigger emotion through channels distinct from vision~\cite{Winters:2014:SES}, though it can compete with data clarity~\cite{Ronnberg:2021:SSP}, an effect also visible in
widely-discussed projects such as Halloran's \textit{The Fallen of
World War II}~\cite{halloran2015fallen}. 
    \item \designIB{Creating cue and atmosphere}. Four participants added surface cues, which can be described as designing for the vibe. P7 incorporated aged paper textures, sepia tones, historical photography, \inlinequote{barbed wire at the bottom,} and smoke around the \inlinequote{1918} section to build the emotional register of the historical period. P12 chose pastel colors, soft oval shapes, and slow animations \inlinequote{like a flower blossoming in time.}
\end{itemize}





\begin{figure}[t]
    \centering
    \includegraphics[width=.9\columnwidth]{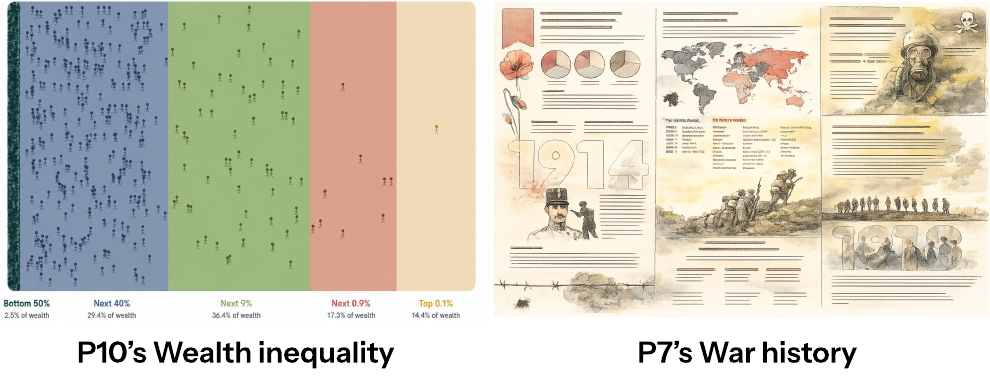} 
    \vspace{-2em}
\end{figure}

    \begin{wrapfigure}[8]{r}{100pt}
    \vspace{-1.5\baselineskip}
    \includegraphics[width=100pt]{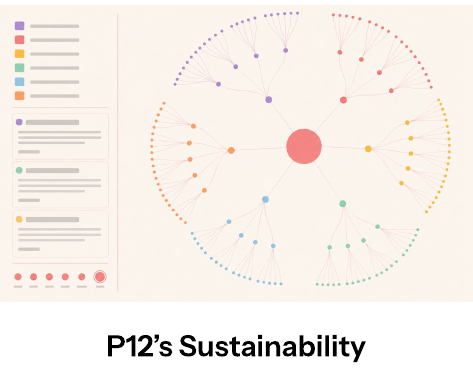} 
    \vspace{-1.5\baselineskip}
    \end{wrapfigure}

\enlargethispage{.75\baselineskip}
    


\paragraph{\designText{Audience as affective calibration.}} Finally, designers consider how different viewers will respond to the data, factoring in audience type, visualization literacy, distribution platform, and cultural background. 
P14 chose a bar chart for scientific audiences who \inlinequote{need precise data,} but a 3D globe for audiences less familiar with standard charts: \inlinequote{Maybe even my parents would have a hard time understanding what a bar chart is. The globe might be easier to understand.} Distribution platforms can shape the visualization design. 
P9 used a scrollytelling website for extended engagement and a slideshow for Instagram. P4 kept the visualization itself aesthetic, letting anger enter through captions on LinkedIn and Instagram. P15 noted that confronting U.S. audiences with climate responsibility can \inlinequote{lead to increased climate denial.} Similarly, Schuster et al.~\cite{Schuster:2026:PPD} found that practitioners designing for broad audiences rely on expertise and platform to characterize their audience.
For this facet, five participants used \designIB{\,their own emotions as inputs to the design process}. P4 had a feedback loop: \inlinequote{...the more emotional I felt, and that emotion informed the data visualization, which in turn informed my emotions.} P15 reflected that this engagement had become \inlinequote{intentional but not necessarily conscious, or it's not conscious anymore.} This aligns with the argument in interaction design that designers' own experience functions as design knowledge~\cite{Zhang:2014:URD} and Lupi's Data Humanism~\cite{Lupi:2017:DHR}, which advocates personal connections with data.\looseness=-10 



\subsection*{Finding 3: Ethical Considerations (RQ3)}
\label{sec:rq3}

Participants also debated ethical reasoning in designing affective visualization, which we categorized into four considerations below.

\paragraph{\ethics{Whether and when emotion is appropriate.}} Five participants offered diverse opinions. P14 allowed more emotion for topics with broad social consensus such as climate change but demanded neutrality for contested ones like elections. P15 noted that the same approach can produce opposite effects across cultures: confronting U.S. audiences with their responsibility for
climate change \inlinequote{actually leads to increased climate denial,} so \inlinequote{the ethical choice is to tamp down the emotions.} Yet P7 had a different position, framing the question as \inlinequote{more of a commercial approach than a moral one,} focused on accessibility rather than moralism.

\paragraph{\ethics{Meaningful engagement vs. manipulation.}} Six participants together offered three criteria for drawing a line between engagement and manipulation. The first criterion was \ethicsUL{intent} and \ethicsUL{self-interest}: whether the designer stands to benefit from shifting the audience's position. P11 argued that \inlinequote{manipulation implies bringing someone to a position that is convenient for you.} P10 framed the same criterion through the lens of responsibility: \inlinequote{if a story or dataset is compelling enough, I don't think you have to be unethical to make somebody feel something.}
The second criterion was \ethicsUL{topic} and \ethicsUL{social consensus}: whether the subject matter carries broad normative agreement. P14 argued that emotional exaggeration is \inlinequote{less bad} for a charity cause than for a political campaign, where values are contested.
The third criterion was \ethicsUL{accuracy}: whether the emotional framing distorts or respects the underlying data. P4 kept emotion out of data processing entirely, \inlinequote{when I'm actually coding and analyzing the data in Python, I am very neutral,} letting emotion enter only at the design stage. 


\paragraph{\ethics{Directing emotion at the right target.}} Three participants also raised the question of \textit{where} emotions should be directed. 
For instance, P15 argued that climate data has \inlinequote{a lot of places for people or big companies to hide and misdirect outrage,} making it important to be \inlinequote{mad at the right things,} and the problem of misdirected outrage (blaming individuals when the causes are structural) is an ethical failure and a failure of causal understanding.\looseness=-10 

\paragraph{\ethics{Balancing emotion with clarity and accuracy.}} Seven participants mentioned that even when emotion was deemed appropriate, they balanced it against clarity and data accuracy. P10 described designing \inlinequote{to evoke some feeling, while making sure it's analytically rigorous and not distorting the data in any way.} P4 resolved this tension by keeping data processing neutral and letting emotion enter through narrative framing such as captions. Despite their disagreements over whether emotion belongs in visualization, all participants agreed that data accuracy was a baseline.

\enlargethispage{\baselineskip}

\section{Discussion \& Conclusion}
\label{sec:discussion}

By interviewing 15 visualization practitioners, we identify three functions that designers intend emotions to serve in
their work, three facets of how
designers work with emotions, and ethical considerations that arise throughout the design process.
In this section, we discuss observations beyond our research questions and future opportunities. 



\paragraph{Affective intent is not always planned.}
Affective intent often takes form during the design process rather than being planned from the start. Only four participants stated that they \textit{\textbf{planned}} their affective goals at the outset (\eg urgency, empathy). The remaining participants described two other modes of affective intent.   
{\bf{\textit{Emergent intent}}} is discovered through the act of designing. P4 began without a clear goal for emotion but found one through an iterative feedback loop between personal emotion and design decisions. 
{\bf{\textit{Tacit intent}}} is even harder to recognize: P15 described their  intent as \inlinequote{not conscious anymore} after years of practice. 

These three modes may not fit within existing visualization design frameworks~\cite{Munzner:2009:NMV, Sedlmair:2012:DSM, McKenna:2014:DAF}, which
describe design as a sequence of cognitive decisions about data, encoding, and analysis. 
They do not yet accommodate decisions about emotion that may be set in advance, emerge through iteration, or remain tacit~\cite{Schon:1983:RPH}. 
The co-evolution of affective intent and design better resonates with what Parsons and Shukla~\cite{Parsons:2026:BPS} recently observed in general visualization design, where problem framing and solution development advance together rather than sequentially.

\paragraph{Evaluation as a key  challenge.}
Evaluation remains a key challenge across our interviews, particularly the gap between designers' affective intent and how that intent is received by viewers. 
P14 described having \inlinequote{no time or budget for user testing,} while P11 worked from \inlinequote{only my understanding} rather than shared evidence. 
Beyond logistics, designers \inlinequote{had no systematic way to assess whether their affective intent was actually received by viewers} (\px{15}). 
Existing instruments mostly capture either viewers' emotional responses or their downstream behaviors. 
The SAM scale~\cite{Blair:2024:QER} measures arousal and valence, while affective learning objectives and their assessments~\cite{LeeRobbins:2022:ALO, LeeRobbins:2026:AAR} target belief change, behavioral intent, or donation. 
These instruments work well when the designer's intent maps onto the metric. But for projects where the goal is to draw attention or sustain engagement, these instruments are not enough. Many projects may not have a clear goal for the outcome. 
For these projects, measuring belief change or behavioral intent would miss what the designer is trying to do. Even for outcome-oriented projects, an outcome measurement does not indicate whether viewers did not notice the visualization, or engage, feel, or act. 
Finally, since designers' affective intent can be emergent or tacit, evaluation may benefit from methods that let designers describe their intent after the fact. 
Design diaries kept during projects can record design decisions about emotion as they form, and retrospective interviews can surface tacit reasoning that designers no longer consciously articulate. 
Recent work in general visualization design~\cite{Parsons:2026:BPS} used such methods to study how problem framing co-evolves with solution development.
Pairing designers' accounts with viewer responses may also surface where designer intent and viewer experience diverge.

\paragraph{Affective impact accumulates across design choices.} Designers did not rely on single visual elements to evoke emotion. P11 combined a knot metaphor drawn from victims' language, individual stories instead of aggregate statistics, and audio readings of excerpts. Other projects similarly combined metaphors, colors, typography, textures, and cultural symbols. Beyond our participants' projects, Periscopic's \textit{U.S. Gun Deaths}~\cite{Periscopic:2013:Guns} layers individual named victims, orange life lines that turn grey at the moment of death, and timing slow enough to prevent skipping.
The impact of emotion in these projects emerged from the accumulation of multiple design choices.
This may help explain why prior experiments testing single-element effects, such as replacing abstract marks with anthropomorphic icons~\cite{Boy:2017:SPB, Morais:2021:CAP}, found limited effects on empathy and prosocial behavior. 
If affective response depends on combined strategies across data, design, and audience facets (\cref{sec:rq2}), then experiments isolating one design choice will underestimate the cumulative effects. Existing frameworks for affective visualization, such as Lan et al.'s~\cite{Lan:2024:AVD} sensation, narrative, behavior, and context categories, treat these as separable strategy types. Our findings suggest these categories overlap and reinforce each other in practice. Therefore, future evaluation work would need to study them in combination rather than in isolation.

\paragraph{Limitations \& Opportunities.}  Finally, we want to note some limitations and opportunities for future work. Our recruitment targeted designers whose work appeared to engage with emotion. Although 15 participants provided sufficient depth for qualitative thematic analysis~\cite{Caine:2016:LSS, Braun:2006:UTA}, it would be informative to extend this sample to contexts such as business intelligence, scientific visualization, or data journalism, where different professional constraints apply, and to client-driven instead of self-initiated projects. Most projects were also static, leaving other strategies such as animation for future work. Our findings are based on self-reports of past design processes, which may present decisions as more intentional than they were in practice. 
Future work can complement interviews with design diaries or think-aloud protocols, and user studies comparing designer intent with viewer experience.
\looseness=-10








\acknowledgments{%
The authors thank Hernisa Kacorri, Taehyun Yang, Zhongzheng Xu, Xiaoyu Liu, Leo Liu, David Guan, Christy Chiu, and Kazi Tasnim Zinat for their valuable feedback, participants for their time, and reviewers for their comments. 
\paragraph{CRediT.} YB: data, methodology, analysis, draft, review, conceptualization. ZW: analysis. KW: review, conceptualization. FY: methodology, draft, review, conceptualization, funds, supervision.  
}

\emergencystretch=3em
\let\oldthebibliography\thebibliography
\renewcommand{\thebibliography}[1]{%
  \oldthebibliography{#1}%
  \setlength{\itemsep}{-1.5pt}%
  \setlength{\parsep}{0pt}%
  \setlength{\parskip}{0pt}%
}

\bibliographystyle{abbrv-doi}

\bibliography{template}

@article{Russell:1980:CMA,
  author = {James A. Russell},
  title = {A Circumplex Model of Affect},
  journal = {Journal of Personality and Social Psychology},
  volume = {39},
  number = {6},
  pages = {1161--1178},
  year = {1980},
  doi = {10.1037/h0077714}
}

@book{Damasio:1994:DES,
  author = {Antonio R. Damasio},
  title = {Descartes' Error: Emotion, Reason, and the Human Brain},
  publisher = {G.P. Putnam's Sons},
  address = {New York},
  year = {1994}
}

@article{Blair:2024:QER,
  author = {Carter Blair and Xiyao Wang and Charles Perin},
  title = {Quantifying Emotional Responses to Immutable Data Characteristics and Designer Choices in Data Visualizations},
  journal = {TVCG},
  volume = {31},
  number = {1},
  pages = {1006--1016},
  year = {2025},
  doi = {10.1109/TVCG.2024.3456361}
}

@article{Lan:2024:AVD,
  author = {Xingyu Lan and Yanqiu Wu and Nan Cao},
  title = {Affective Visualization Design: Leveraging the Emotional Impact of Data},
  journal = {TVCG},
  volume = {30},
  number = {1},
  pages = {1--11},
  year = {2024},
  doi = {10.1109/TVCG.2023.3327385}
}

@article{Prantl:2024:URE,
  author = {Verena Ingrid Prantl and Torsten M{\"o}ller and Laura Koesten},
  title = {Untangling Rhetoric, Pathos, and Aesthetics in Data Visualization},
  journal = {TVCG},
  volume = {32},
  number = {2},
  pages = {2435--2453},
  year = {2026},
  doi = {10.1109/TVCG.2025.3628181}
}

@inproceedings{Bartram:2017:ACV,
  author = {Lyn Bartram and Abhisekh Patra and Maureen Stone},
  title = {Affective Color in Visualization},
  booktitle = {CHI},
  year = {2017},
  doi = {10.1145/3025453.3026041}
}

@inproceedings{Lan:2022:NEP,
  author = {Xingyu Lan and Yanqiu Wu and Yang Shi and Qing Chen and Nan Cao},
  title = {Negative Emotions, Positive Outcomes? {E}xploring the Communication of Negativity in Serious Data Stories},
  booktitle = {CHI},
  year = {2022},
  doi = {10.1145/3491102.3517530}
}

@inproceedings{Boy:2017:SPB,
  author = {Jeremy Boy and Anshul Vikram Pandey and John Emerson and Margaret Satterthwaite and Oded Nov and Enrico Bertini},
  title = {Showing People Behind Data: Does Anthropomorphizing Visualizations Elicit More Empathy for Human Rights Data?},
  booktitle = {CHI},
  year = {2017},
  doi = {10.1145/3025453.3025512}
}

@inproceedings{Morais:2021:CAP,
  author = {Luiz Morais and Yvonne Jansen and Nazareno Andrade and Pierre Dragicevic},
  title = {Can Anthropographics Promote Prosociality? {A} Review and Large-Sample Study},
  booktitle = {CHI},
  year = {2021},
  doi = {10.1145/3411764.3445637}
}

@article{Slovic:2007:ILM,
  author = {Paul Slovic},
  title = {``If {I} Look at the Mass {I} Will Never Act'': {P}sychic Numbing and Genocide},
  journal = {Judgment and Decision Making},
  volume = {2},
  number = {2},
  pages = {79--95},
  year = {2007},
  doi = {10.1017/S1930297500000061}
}

@article{Parsons:2022:UDV,
  author = {Paul Parsons},
  title = {Understanding Data Visualization Design Practice},
  journal = {TVCG},
  volume = {28},
  number = {1},
  pages = {665--675},
  year = {2022},
  doi = {10.1109/TVCG.2021.3114959}
}

@inproceedings{Baigelenov:2025:HVD,
  author = {Ali Baigelenov and Prakash Shukla and Paul Parsons},
  title = {How Visualization Designers Perceive and Use Inspiration},
  booktitle = {CHI},
  publisher = {ACM},
  address = {New York},
  year = {2025},
  doi = {10.1145/3706598.3714191}
}

@article{Zhang:2023:VDP,
  author = {Yixuan Zhang and Yifan Sun and Joseph D. Gaggiano and Neha Kumar and Clio Andris and Andrea G. Parker},
  title = {Visualization Design Practices in a Crisis: Behind the Scenes with {COVID}-19 Dashboard Creators},
  journal = {TVCG},
  volume = {29},
  number = {1},
  pages = {1037--1047},
  year = {2023},
  doi = {10.1109/TVCG.2022.3209493}
}

@article{Vastfjall:2014:CFA,
  author = {Daniel V{\"a}stfj{\"a}ll and Paul Slovic and Marcus Mayorga and Ellen Peters},
  title = {Compassion Fade: Affect and Charity Are Greatest for a Single Child in Need},
  journal = {PLOS ONE},
  volume = {9},
  number = {6},
  pages = {e100115},
  year = {2014},
  doi = {10.1371/journal.pone.0100115}
}

@article{Segel:2010:NVT,
  author = {Edward Segel and Jeffrey Heer},
  title = {Narrative Visualization: Telling Stories with Data},
  journal = {TVCG},
  volume = {16},
  number = {6},
  pages = {1139--1148},
  year = {2010},
  doi = {10.1109/TVCG.2010.179}
}

@misc{Lupi:2017:DHR,
  author = {Giorgia Lupi},
  title = {Data Humanism: The Revolutionary Future of Data Visualization},
  year = {2017},
  howpublished = {http://giorgialupi.com/data-humanism-my-manifesto-for-a-new-data-wold},
  note = {Accessed: 2026-06-27}
}

@article{LeeRobbins:2022:ALO,
  author = {Elsie Lee-Robbins and Eytan Adar},
  title = {Affective Learning Objectives for Communicative Visualizations},
  journal = {TVCG},
  volume = {29},
  number = {1},
  pages = {1--11},
  year = {2023},
  doi = {10.1109/TVCG.2022.3209500}
}

@inproceedings{Zhang:2014:URD,
  author = {Xiao Zhang and Ron Wakkary},
  title = {Understanding the Role of Designers' Personal Experiences in Interaction Design Practice},
  booktitle = {DIS},
  year = {2014},
  doi = {10.1145/2598510.2598556}
}

@inproceedings{Caine:2016:LSS,
  author = {Kelly Caine},
  title = {Local Standards for Sample Size at {CHI}},
  booktitle = {CHI},
  pages = {981--992},
  year = {2016},
  doi = {10.1145/2858036.2858498}
}

@article{Fereday:2006:DRU,
  author = {Jennifer Fereday and Eimear Muir-Cochrane},
  title = {Demonstrating Rigor Using Thematic Analysis: {A} Hybrid Approach of Inductive and Deductive Coding and Theme Development},
  journal = {International Journal of Qualitative Methods},
  volume = {5},
  number = {1},
  pages = {80--92},
  year = {2006},
  doi = {10.1177/160940690600500107}
}

@book{Swain:2018:HAT,
  author = {Jon Swain},
  title = {A Hybrid Approach to Thematic Analysis in Qualitative Research: Using a Practical Example},
  publisher = {SAGE Publications},
  address = {London},
  year = {2018},
  doi = {10.4135/9781526435477}
}

@article{Munzner:2009:NMV,
  author = {Tamara Munzner},
  title = {A Nested Model for Visualization Design and Validation},
  journal = {TVCG},
  volume = {15},
  number = {6},
  pages = {921--928},
  year = {2009},
  doi = {10.1109/TVCG.2009.111}
}

@article{Sedlmair:2012:DSM,
  author = {Michael Sedlmair and Miriah Meyer and Tamara Munzner},
  title = {Design Study Methodology: Reflections from the Trenches and the Stacks},
  journal = {TVCG},
  volume = {18},
  number = {12},
  pages = {2431--2440},
  year = {2012},
  doi = {10.1109/TVCG.2012.213}
}

@article{McKenna:2014:DAF,
  author = {Sean McKenna and Dominika Mazur and James Agutter and Miriah Meyer},
  title = {Design Activity Framework for Visualization Design},
  journal = {TVCG},
  volume = {20},
  number = {12},
  pages = {2191--2200},
  year = {2014},
  doi = {10.1109/TVCG.2014.2346331}
}

@article{Nabi:2003:EFE,
  author = {Robin L. Nabi},
  title = {Exploring the Framing Effects of Emotion: {D}o Discrete Emotions Differentially Influence Information Accessibility, Information Seeking, and Policy Preference?},
  journal = {Communication Research},
  volume = {30},
  number = {2},
  pages = {224--247},
  year = {2003},
  doi = {10.1177/0093650202250881}
}

@article{LeeRobbins:2026:AAR,
  author = {Elsie Lee-Robbins and Eytan Adar},
  title = {Assessing Affective Objectives for Communicative Visualizations},
  journal = {arXiv:2604.01183},
  year = {2026}
}

@article{Winters:2014:SES,
  author = {R. Michael Winters and Marcelo M. Wanderley},
  title = {Sonification of Emotion: Strategies and Results from the Intersection with Music},
  journal = {Organised Sound},
  volume = {19},
  number = {1},
  pages = {60--69},
  year = {2014},
  doi = {10.1017/S1355771813000411}
}

@article{Ronnberg:2021:SSP,
  author = {Niklas R{\"o}nnberg},
  title = {Sonification Supports Perception of Brightness Contrast},
  journal = {Journal on Multimodal User Interfaces},
  volume = {15},
  pages = {171--178},
  year = {2021},
  doi = {10.1007/s12193-020-00378-1}
}

@book{Schon:1983:RPH,
  author = {Donald A. Sch{\"o}n},
  title = {The Reflective Practitioner: How Professionals Think in Action},
  publisher = {Basic Books},
  address = {New York},
  year = {1983}
}

@article{Nabi:2018:FCC,
  author = {Robin L. Nabi and Abel Gustafson and Risa Jensen},
  title = {Framing Climate Change: Exploring the Role of Emotion in Generating Advocacy Behavior},
  journal = {Science Communication},
  volume = {40},
  number = {4},
  pages = {442--468},
  year = {2018},
  doi = {10.1177/1075547018776019}
}

@article{Braun:2006:UTA,
  author = {Virginia Braun and Victoria Clarke},
  title = {Using Thematic Analysis in Psychology},
  journal = {Qualitative Research in Psychology},
  volume = {3},
  number = {2},
  pages = {77--101},
  year = {2006},
  doi = {10.1191/1478088706qp063oa}
}

@article{Parsons:2026:BPS,
  author = {Paul C. Parsons and Prakash Chandra Shukla},
  title = {Beyond Problem Solving: Framing and Problem--Solution Co-Evolution in Data Visualization Design},
  journal = {TVCG},
  volume = {32},
  number = {1},
  pages = {24--34},
  year = {2026},
  doi = {10.1109/TVCG.2025.3633866}
}

@article{Lan:2025:MWF,
  author = {Xingyu Lan and Yutong Yang and Yifan Wang},
  title = {``{M}apping What {I} Feel'': Understanding Affective Geovisualization Design Through the Lens of People--Place Relationships},
  journal = {TVCG},
  volume = {32},
  number = {1},
  pages = {145--155},
  year = {2026},
  doi = {10.1109/TVCG.2025.3633878}
}

@book{Card:1999:RIV,
  author = {Stuart K. Card and Jock D. Mackinlay and Ben Shneiderman},
  title = {Readings in Information Visualization: Using Vision to Think},
  publisher = {Morgan Kaufmann},
  address = {San Francisco},
  year = {1999}
}

@book{Munzner:2014:VAD,
  author = {Tamara Munzner},
  title = {Visualization Analysis and Design},
  publisher = {CRC Press},
  address = {Boca Raton, FL},
  year = {2014}
}

@misc{InfoBea,
  title        = {{Information Is Beautiful}},
  year         = {2019},
  howpublished = {https://informationisbeautiful.net/wdvp/gallery-2019},
  note = {Accessed: 2026-06-27}
}

@misc{Scarr:2011:Iraq,
  author       = {Simon Scarr},
  title        = {{Iraq's Bloody Toll}},
  howpublished = {South China Morning Post},
  year         = {2011},
  url          = {https://www.simonscarr.com/iraqs-bloody-toll},
  note = {Accessed: 2026-06-27}
}

@article{Yang:2024:SwayPublic,
  author = {Fumeng Yang and Mandi Cai and Chloe Mortenson and Hoda Fakhari and Ayse D. Lokmanoglu and Nicholas Diakopoulos and Erik C. Nisbet and Matthew Kay},
  title = {Swaying the Public? {I}mpacts of Election Forecast Visualizations on Emotion, Trust, and Intention in the 2022 {U.S.} Midterms},
  journal = {TVCG},
  volume = {30},
  number = {1},
  pages = {23--33},
  year = {2024},
  doi = {10.1109/TVCG.2023.3327356}
}

@misc{Periscopic:2013:Guns,
  author = {{Periscopic}},
  title = {{U.S. Gun Deaths}},
  year = {2013},
  howpublished = {\url{https://guns.periscopic.com/}},
  note = {Interactive data visualization, Accessed: 2026-06-27}
}

@misc{halloran2015fallen,
  author = {Halloran, N.},
  title = {The {F}allen of {W}orld {W}ar {II}},
  year = {2015},
  howpublished = {\url{http://www.fallen.io/ww2/}},
  note = {Interactive data-driven documentary}
}

@inproceedings{Alhazwani:2025:DHD,
  author = {Ibrahim Al-Hazwani and Ke Er Amy Zhang and Laura Garrison and J{\"u}rgen Bernard},
  title = {Data Humanism Decoded: {A} Characterization of Its Principles to Bridge Data Visualization Researchers and Practitioners},
  booktitle = {IEEE VIS Short Papers},
  year = {2025},
  doi = {10.1109/VIS60296.2025.00057}
}

@article{Graf:2015:DPP,
  author  = {Laura K. M. Graf and Jan R. Landwehr},
  title   = {A Dual-Process Perspective on Fluency-Based Aesthetics: {T}he Pleasure-Interest Model of Aesthetic Liking},
  journal = {Personality and Social Psychology Review},
  volume  = {19},
  number  = {4},
  pages   = {395--410},
  year    = {2015},
  doi     = {10.1177/1088868315574978}
}

@inproceedings{Schuster:2026:PPD,
  author = {Regina Schuster and Kathleen Gregory and Torsten M{\"o}ller and Laura Koesten},
  title = {Practitioners' Perspectives on Designing Data Visualizations for the General Public},
  booktitle = {CHI},
  pages = {1--19},
  year = {2026},
  doi = {10.1145/3772318.3790627}
}

@article{Kale:2020:VRS,
  author  = {Alex Kale and Matthew Kay and Jessica Hullman},
  title   = {Visual Reasoning Strategies for Effect Size Judgments and Decisions},
  journal = {TVCG},
  volume  = {27},
  number  = {2},
  pages   = {272--282},
  year    = {2021},
  doi     = {10.1109/TVCG.2020.3030335}
}
\end{document}